# Generalization of Kirchhoff's Law: The inherent relations between quantum efficiency and emissivity

**M. KURTULIK[1][†], M. SHIMANOVICH[2][†], R. WEILL[2], A. MANOR[1], M. SHUSTOV[2], AND C. ROTSCHILD[1,2,*]**

[1]*Russell Berrie Nanotechnology Institute, Technion − Israel Institute of Technology, Haifa 3200003, Israel*

[2]*Department of Mechanical Engineering, Technion − Israel Institute of Technology, Haifa 3200003, Israel*

[†] *Equal contribution as first author*

*\*Corresponding author: carmelr@technion.ac.il*

**Abstract:** Planck's law of thermal radiation depends on the temperature, $T$, and the emissivity, $\varepsilon$, of a body, where emissivity is the coupling of heat to radiation that depends on both phonon-electron nonradiative interactions and electron-photon radiative interactions. Another property of a body is absorptivity, $\alpha$, which only depends on the electron-photon radiative interactions. At thermodynamic equilibrium, nonradiative interactions are balanced, resulting in Kirchhoff's law of thermal radiation that equals these two properties, i.e., $\varepsilon = \alpha$. For non-equilibrium, Quantum efficiency (QE) describes the statistics of photon emission, which like emissivity depends on both radiative and nonradiative interactions. Past generalized Planck's equation extends Kirchhoff's law out of equilibrium by scaling the emissivity with the pump-dependent chemical-potential $\mu$, obscuring the relations between the body properties. Here we theoretically and experimentally demonstrate a prime equation relating these properties in the form of $\varepsilon = \alpha(1 - QE)$, which is in agreement with a recent universal modal radiation laws for all thermal emitters. At equilibrium, these relations are reduced to Kirchhoff's law. Our work lays out the fundamental evolution of non-thermal emission with temperature, which is critical for the development of lighting and energy devices.

**Introduction**

Planck thermal radiation is the emission of a body at non-zero temperature. It is characterized by the temperature of the body $T$, and its ability to thermally emit radiation, given by emissivity $\varepsilon$, which is a material and cavity property reflecting the contribution of both radiative $\gamma_r$, and non-radiative $\gamma_{nr}$ rates. While $\gamma_{nr}$ couples heat to excite electrons, $\gamma_r$ couples the excited electrons to photon emission. In the reverse process, these rates respectively govern non-radiative recombination and photon absorption, denoted by $\alpha$. Although absorptivity depends only on the radiative rates, while emissivity depends on both radiative and non-radiative rates, at thermodynamic equilibrium, thermal excitation is balanced by the non-radiative recombinations, leading to the equality between emissivity and absorptivity $\varepsilon = \alpha$, known as Kirchhoff's law for thermal emission. Out of equilibrium, i.e., a body emits radiation into an environment having a different temperature or chemical potential, Kirchhoff's law does not apply, since non-radiative processes are not canceled out (thermal excitation ≠ non-radiative recombination). This phenomenon can be seen in semiconductors, light-emitting diodes, and, in atoms, in photoluminescence [1]. An extreme example of non-equilibrium is when absorptivity becomes negative at high excitation rates while the emissivity is positive [2-4].

Photoluminescence (PL), as an example of non-equilibrium emission, involves the absorption of photons followed by fast thermalization of the excited electrons and emission of (typically low-energy, red-shifted) photons [5-11]. Since thermalization results in the energy difference between the incoming and outgoing photon energy, it is defined as out of equilibrium. Nevertheless, thermodynamics statistically analyzes PL at quasi-equilibrium [12-15], using the conventional thermodynamic variables such as temperature, emissivity, and chemical potential µ [16, 17], where the latter describes the excitation above thermal. The generalized Planck's formula describes such non-equilibrium emission [17, 18] as:

$$L(h\nu, T, \mu) = \varepsilon(h\nu) \frac{(2\nu^2)}{c^3} \frac{1}{e^{\frac{h\nu-\mu}{k_B T}} - 1} \cong L_{Th}(h\nu, T) e^{\frac{\mu}{k_B T}} \quad (1)$$

where $L$ is the spectral radiance (having units of watts per frequency, per solid angle, and per unit area); ε is the emissivity, which is a material and cavity property that depends on the density of states (DoS) of the radiative and non-radiative transitions; $T$ is the temperature; $h\nu$ is the photon energy; $k_B$ is Boltzmann's constant; and $L_{Th}$ is the thermal radiance when the pump is off. The chemical potential $\mu$ is the Gibbs free energy per emitted photon and is also the gap that is opened in semiconductors between the quasi-fermi levels under excitation. Eq. 1 is relevant at a specific frequency band, where the chemical potential is a constant. Without knowing the inherent relation of $\mu(T)$, however, Eq. 1 can only be fitted to an observed PL, without predicting its evolution. Previous studies on non-equilibrium emission by external pump excitation assume the temperature of the surroundings to be the same as the PL body [1-4]. In such cases, Eq. 1 can be seen as a generalization of Kirchhoff's law, demonstrating that the line shape of the non-thermal emission is the same as the thermal emission enhanced by the (scalar) chemical potential, $\varepsilon(h\nu) e^{\frac{\mu}{k_B T}}$ [13]. In such a formalism, however, the emissivity becomes pump-dependent, losing its original meaning as a material and cavity property, and unable to provide insight into the temperature-dependent evolution of the radiation. This state of affairs leaves many open questions. For example, what is the inherent relation between emissivity and absorptivity for a material that thermally emits ($T > 0$) toward an environment at a different temperature? Is there an inherent relation between emissivity and QE, since both depend on the same parameters? In similar to Kirchhoff's law, what is the inherent evolution of non-equilibrium radiation with temperature? The aim of this paper is to extend Kirchhoff's law as a material and cavity property to non-equilibrium and to define the inherent relations between fundamental properties of radiation.

**METHODS**

Under pump excitation, Eq. 1 also defines the PL quantum efficiency (QE) as the ratio between the emitted and absorbed photon rates:

$$QE = \frac{\#\ of\ emitted\ photons}{\#\ of\ absorbed\ photons} = \frac{L(T=0,\mu)}{\alpha \cdot L_{pump}} \qquad (2)$$

where $L(T=0,\mu)$ is the rate of photons emitted by a pumped body at $T=0$, and $L_{pump}(T_p)$ is the incoming photon rate into the system with a brightness temperature $T_p$, which defines the radiance of the pump, at a specific wavelength, as equal to black body radiance at the temperature $T_p$. QE is a material and geometric property of the body defined at low temperatures where thermal excitation is negligible and, similarly to emissivity, it reflects the competition between radiative and nonradiative rates, where higher $\gamma_r$ and lower $\gamma_{nr}$ support higher QE.

Recent experimental work on PL at elevated temperatures [19], shows a quasi-conserved photon emission rate in the low-temperature range, far above thermal emission, which corresponds to the QE. At this temperature range, the PL spectrum is blue-shifted with temperature rise. At higher temperatures, the PL rate increases exponentially, approaching the thermal emission. Figure 1 illustrates the PL rate evolution with temperature based on [19], excluding the very high-temperature range (above 1500K), which will be discussed at the end of the paper. The green line depicts the pump photon rate $L_p$, while the red line depicts the absorbed photon rate $\alpha \times L_p$, where α is the absorptivity. The blue curves are the PL (solid line) and thermal emission (dashed line) for a specific material and geometry for pump-on and pump-off respectively.

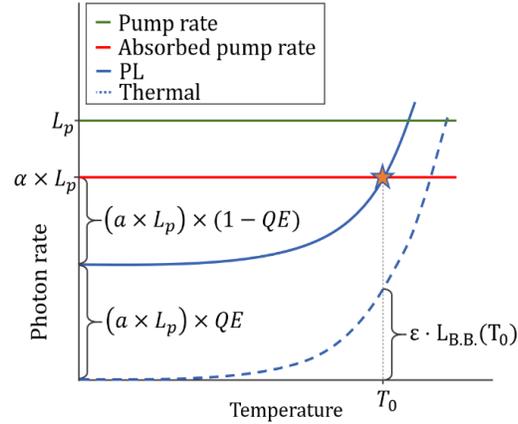

Fig. 1. Illustration of PL vs. temperature.

At low temperature, the PL rate is defined by $\alpha \times L_p \times QE$. As the temperature increases, the PL increases monotonically up to a temperature at which the PL photon rate equals the absorbed photon rate (marked with a star in figure 1), defined as the universal point. The portion of electrons that are photonically excited and recombine non-radiatively at low temperature is $(1 - QE) \times \alpha \times L_p$, where at the universal point, this portion is restored by thermal contribution. We hypothesize that this portion equals the thermal emission of a material such that:

$$\varepsilon \times L_{B.B.}(h\nu, T_0) = (1 - QE) \times \alpha \times L_p \qquad (3)$$

$L_{B.B.}(h\nu, T_0)$ is the black body emission at $T_0$, which, together with $L_p$, are quantities independent of the specific body properties, while $\varepsilon, \alpha$, and the QE are material and geometry properties. Thus, we can deduct from this equality the following relations:

$$\begin{cases} (i) & L_{B.B.}(h\nu, T_0) = L_p \\ (ii) & \varepsilon = \alpha \times (1 - QE)) \end{cases} \qquad (4)$$

By definition, $T_0$ in Eq. 4 (i) is the brightness temperature of the pump, manifesting the zeroth law of thermodynamics at the universal point, where the brightness temperature of the pump equals the body temperature.

Intuitively, in light of the 2nd law of thermodynamics, which states that heat flows from hot to cold bodies, and thus the temperature of a heated body cannot exceed the temperature of the excitation source and, similarly, the radiance of a body cannot exceed the radiance of the pump source at temperatures below the critical temperature. Therefore, according to figure 1, ε must be reduced, compared to α, by the QE amount, as in Eq. 4(ii); $\varepsilon = \alpha(1 - QE)$. That is, for high

QE materials, the thermal evolution is weak compared to its absorptivity and vice versa. This also agrees with the understanding that high-QE materials have low phonon-electron nonradiative rates.

Recent theoretical work on a universal modal radiation law [20] supports a very similar conclusion while analyzing a scattering system in thermodynamic equilibrium. We treat photoluminescence as part of the scattering coefficient $s_p^2$ and reach our prime equation. The similarity between the two approaches is explained in greater detail in supplementary material, section 1. In the following, we first prove our hypothesis numerically and then continue with experimental validation.

**RESULTS AND DISCUSSION**

Following Siegman [21], we study the case of a three-level system in a cavity emitting toward a zero-K environment (describing, for example, the 830nm–900nm emission lines of $Nd^{+3}$ as depicted in [19]). Such a detailed balance considers only photonic, electronic, and phononic transitions and omits other processes associated, for example, with defects, which may cause additional temperature-dependent quenching. As such, this model describes the upper limit of temperature-dependent luminescence. Nevertheless, any additional factors can be embedded in the radiative and non-radiative rates for a specific solution.

Figure 2a shows the considered energy levels having a ground state and a broad excited level consisting of two closely spaced levels, with very fast non-radiative thermalization between them ($\gamma_{nr23}$). This drives towards a Boltzmann distribution of the excited electron populations between the two levels, $n_2$ and $n_3$ [22] (see Section 2 in the supplementary material). Also, here, the brightness temperature $T_p$ defines the rate within the angular and spectral coupling window $\Gamma_p$. The system is described by:

$$\frac{dn_2}{dt} = (n_1 - n_2)B_{r12}n_{ph12} - n_2\gamma_{r12} + (n_1 - n_2)B_{nr12}n_{pn12} - \quad (5a)$$

$$n_2\gamma_{nr12} + (n_3 - n_2)B_{nr23}n_{pn23} + n_3\gamma_{nr23}$$

$$\frac{dn_3}{dt} = (n_1 - n_3)B_{r13}n_{ph13} - n_3\gamma_{r13} + (n_1 - n_3)B_{nr13}n_{pn13} - \quad (5b)$$

$$n_3\gamma_{nr13} - (n_3 - n_2)B_{nr23}n_{pn23} - n_3\gamma_{nr23}$$

$$4\pi \cdot \Delta\nu \cdot \frac{dn_{ph12}}{dt} = n_{pump12}\Gamma_p c - n_{ph12}\Gamma_{12}c - (n_1 - \quad (5c)$$

$$n_2)B_{r12}n_{ph12} + n_2\gamma_{r12}$$

$$4\pi \cdot \Delta\nu \cdot \frac{dn_{ph13}}{dt} = n_{pump13}\Gamma_p c - n_{ph13}\Gamma_{13}c - (n_1 - \quad (5d)$$

$$n_3)B_{r13}n_{ph13} + n_3\gamma_{r13}$$

where $n_1, n_2, n_3$ are the electron population densities of the ground and excited states, respectively, satisfying $N = n_1 + n_2 + n_3$, where $N$ is the total density of atoms in the system; $c$ is the speed of light; $\gamma_{r12}, \gamma_{r13}$ are the radiative spontaneous rates; $\gamma_{nr12}, \gamma_{nr13}$ are the nonradiative spontaneous rates, from both excited levels to the ground state, with units of [1/s]; $\gamma_{nr23} \gg \gamma_{r12}, \gamma_{r13}, \gamma_{nr12}, \gamma_{nr13}$ is the nonradiative rate between the excited energy levels; $\Gamma_p$, $\Gamma_{12}, \Gamma_{13}$ are the coupling coefficients in and out of the cavity, respectively, having units of $\left[\frac{\Delta\nu \cdot Sr}{m}\right]$; $n_{ph12}, n_{ph13}$ are the radiation field densities inside the cavity, having units of $[\#/(\Delta\nu \cdot Sr \cdot m^3)]$, while $n_{pump12} = DoS_{ph}\left[\exp\left(\frac{E_{12}}{kT_p}\right) - 1\right]^{-1}$ and $n_{pump13} = DoS_{ph}\left[\exp\left(\frac{E_{13}}{kT_p}\right) - 1\right]^{-1}$ are the optical fields induced by the pump having a black-body distribution according to the brightness temperature $T_p$. Their sum equals $L_p$. Further, $B_{r12} = \frac{\gamma_{r12}}{DoS_{ph}(E_{12})}$, $B_{r13} = \frac{\gamma_{r13}}{DoS_{ph}(E_{13})}$, and $B_{nr12} = \frac{\gamma_{nr12}}{DoS_{pn}(E_{12})}$, and $B_{nr13} = \frac{\gamma_{nr13}}{DoS_{pn}(E_{13})}$ are the Einstein coefficients for the stimulated absorption or emission rates for both radiative and non-radiative processes [12]. Under fast thermalization, phonons obey the equilibrium distribution, which is given by

$$n_{pn12} = DoS_{pn}(E_{12})\left[exp\left(\frac{E_{12}}{kT}\right) - 1\right]^{-1}, n_{pn13} = DoS_{pn}(E_{13})\left[exp\left(\frac{E_{13}}{kT}\right) - 1\right]^{-1}$$

and $n_{pn23} = DoS_{pn}(E_{23})\left[exp\left(\frac{E_{23}}{kT}\right) - 1\right]^{-1}$ [23]. In this formalism, $DoS_{pn}$ and $DoS_{ph}$ are the corresponding densities of states or the phonons and photons, respectively. Solving the equations for different regimes reveals different observables. For simplicity, we plot the solution for various values of $\gamma_r$ and $\gamma_{nr}$ as temperature-independent. For the temperature-dependent case, the solution can be depicted as crossing between different curves at different temperatures.

The solutions for the photon rate (radiative transitions to the ground state), given by $n_{ph12}\Gamma_{12}$ and $n_{ph13}\Gamma_{13}$, for equally incoming and output coupling rates—$\Gamma_{12} = \Gamma_{13} = \Gamma_p = \Gamma$—under optical excitation $n_{pump12}\Gamma_p$ and $n_{pump13}\Gamma_p$ at $T_p = 1000K$ for various QEs (0, 0.5, and 1), are depicted in figure 2b. These three cases all have the same absorptivity, depending on $\gamma_r$, and different losses due to different $\gamma_{nr}$ rates. The red line describes both the absorbed pump rate and QE=1. The black line in figure 2b describes the thermal emission $\varepsilon(\gamma_r) = \alpha(\gamma_r)$, when setting QE=0 (while keeping $\gamma_r$ constant), describing the case where $\gamma_{nr12}, \gamma_{nr13} \gg \gamma_{r12}, \gamma_{r13}$. Under this regime, nearly all the absorbed photons recombine nonradiatively, which results in thermal emission that increases exponentially until it reaches the QE=1 line at the universal point $T_c = T_p$. Such a body with the QE=0, emits the same amount of radiation to the environment regardless of the optical excitation. It is in the local equilibrium set by the dominant nonradiative rates, and it obeys Kirchhoff's law even out of equilibrium, as was shown in [24]. The blue line describes the PL emission of QE=0.5. At 0K, half of the absorbed photons are lost due to nonradiative recombination $\gamma_{nr}$. In a low (non-zero) temperature range, the total photon rate (the sum of $n_{ph12}\Gamma_{12}$ and $n_{ph12}\Gamma_{13}$) is quasi-conserved, accompanied by a blue-shift of the spectrum-emitted photon rate $n_{ph12}\Gamma_{12}$ that decreases while $n_{ph13}\Gamma_{13}$ increases, as the temperature increases. The ratio between these emissions is given by the Boltzmann distribution as long as $\gamma_{nr23} \gg \gamma_{r12}, \gamma_{r13}, \gamma_{nr12}, \gamma_{nr13}$ [22]. A further increase in temperature leads to an increase in the photon rate until it reaches the cross between the QE=1 case (red line) and the QE=0 case (black line) at the universal point $T_c = T_p$. The photon rate continues, thereafter, to rise above it.

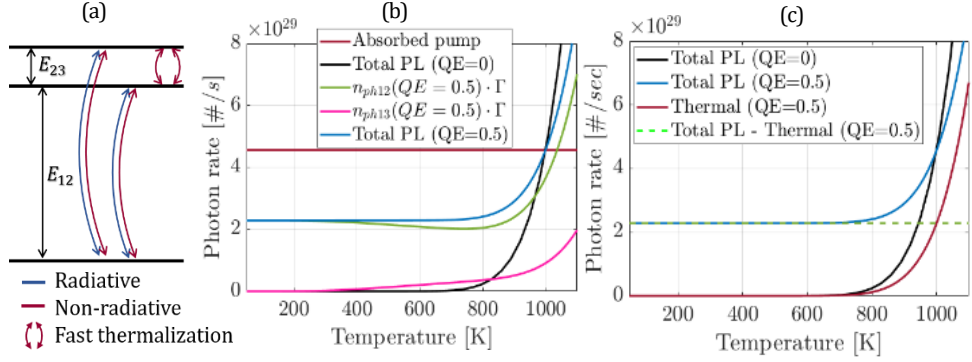

Fig. 2. (a) A three-level system with fast non-radiative thermalization between upper energy levels. (b) PL emission for three different QE: QE=0, 0.5 and 1 systems (black, blue and red lines, respectively), optically excited by a pump at brightness temperature $T_p = 1000K$. The green and purple lines show individual emissions from lower and higher excited energy levels for the QE=0.5 case. (c) Above the critical temperature $T > T_c$, the PL emission (blue line) is bounded by a thermal body when QE=0 (black line) and by the thermal emission for the specific QE (QE=0.5, red line). The dashed green line shows the constant difference between the PL emission and the thermal emission for QE=0.5.

In the absence of an optical pump $n_{pump} = 0$, one gets the thermal emission for the body for any QE emitting toward a zero-K environment. Figure 2c shows the PL line for QE=0.5 (blue line) in addition to its corresponding thermal emission (red line). For comparison, we also show the thermal emission for the case of QE=0 (black line). The difference between the two emission lines, the PL and thermal lines for QE=0.5, is a constant, invariant of temperature (depicted by the green dashed line).

This general solution is, as far as we know, the first explanation for the experimentally observed transition from the rate-conservation region accompanied by a blue shift to thermal emission, where the photon rate increases at any wavelength, as shown in [19]. Evidently, the thermal emission for QE>0 is reduced, compared to the thermal emission of QE=0, due to a lower $\gamma_{nr}$ value compared to $\gamma_r$. In addition, the PL emission beyond the universal point is restricted to remaining between the QE=0 line and the thermal line for the same QE.

We find from figure 2c that the ratio between the thermal emission of a body (for QE>0 in the absence of a pump) and the emission curve of a thermal body having the same absorptivity (QE=0 and $\varepsilon_{QE=0} = \alpha$) is a temperature-independent and QE-dependent constant we name $\varepsilon_{QE>0}$, which results in $\varepsilon_{QE>0} = 1 - QE$. It is evident, that the thermal contribution leads to recovery of the universal point and balancing nonradiative recombination which are proportional to $\alpha \times (1 - QE)$. As suggested by Eq. 4(ii), the total emissivity can be expressed as:

$$\varepsilon = \varepsilon_{QE=0} \times \varepsilon_{QE>0} = \alpha \times (1 - QE) \qquad (6)$$

The total thermal emission of a body toward a zero-K environment (in absence of a pump) is:

$$L(h\nu, T) = \Gamma \times \alpha \times (1 - QE) \times \frac{2\nu^2}{c^3} \frac{1}{e^{\frac{h\nu}{k_B T}} - 1} \qquad (7)$$

For the case where $\gamma_r$ and $\gamma_{nr}$ are temperature dependent, the emission curve crosses from one QE line to the other, yet still reaching the universal point. For this reason, Eq 6. is valid also in the form $\varepsilon(T) = \alpha(T) \times (1 - QE(T))$. For the simple case of a two-level system, the analytical solution for Eq. 6 and the QE as a function of $\gamma_r$ and $\gamma_{nr}$ is presented in Section 3 in the supplementary material.

In addition, the equations show, that in the case when the coupling rate $\Gamma \to 0$, any material approach QE=0. This means that the emission from any material enclosed in an optical cavity becomes Black-body emission. The cavity reduces the photons escaping the system thereby effectively reducing $\gamma_r$, while unchanging $\gamma_{nr}$ which is a material property. The relation $\gamma_{nr} \gg \gamma_r$, together with multiple self-absorption events in the cavity that effectively enhance absorptivity results in a Black-body.

Moving to the experiment, for measuring the emissivity of materials with different QEs, we choose Erbium-doped YAG (Yttrium Aluminum Garnet, $Y_3Al_5O_{12}$) crystals with four different $Er^{+3}$ ion concentrations having mass ratios of 0.2%, 2%, 3% and 5% [23]. The different concentrations affect the QE through self-quenching, with minimal change in the bandgap. The dimensions of the 0.2% and 3% $Er^{+3}$ ion concentration crystals are $10 \times 10 \times 4\ mm^3$ (Roditi), while the 2% and 5% Er:YAG crystals are $11 \times 11 \times 5\ mm^3$ (Crytur). As illustrated in figure 3a, each sample is placed on a silver plate, which is heated by a ceramic heater (Thorlabs HT19R). Due to the high thermal conductivity of silver, along with its relatively low emissivity, it is possible to uniformly heat the sample, as well as perform temperature and thermal emission measurements. The temperature of the silver plate is measured with a k-type thermocouple and controlled with a PID controller (Eurotherm 2416). The thermocouple is placed inside a blind hole, with its tips at the center of the silver plate. The sample is covered with an additional silver plate, having an additional thermocouple to monitor and minimize the temperature gradient over the sample. The power spectrum is acquired using multimode fiber, coupled to a calibrated spectrometer (Cornerstore 260) connected to an InGaAs detector (Andor iDus), placed in the vicinity of the sample face. The acquired power spectrum is normalized by the emission area and solid angle.

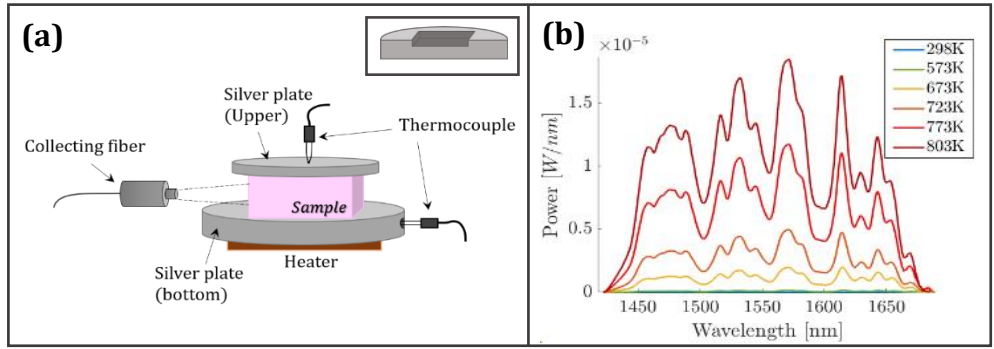

Fig. 3. Experimental setup and measured power spectra of the thermal emission. (a) The experimental setup. Each Er:YAG sample is placed on a silver plate that is positioned on a heater. The lower silver plate is grooved to align the sample (shown in the inset). The sample is heated by a PID-controlled heater. The temperature of the lower plate is measured by a thermocouple placed in a blind hole, drilled in the silver plate. The emitted thermal radiation of the sample is collected by a fiber and the power spectrum is measured by a calibrated spectrometer. To increase emission into the peripheral directions, the sample is covered on top with a second (non-heated)silver plate. (b) The power spectrum of the thermal emission of the 5% Er:YAG crystal.

As an example, the thermal emission of the 5% Er:YAG crystal at the temperature range of 298-803K is presented in figure 3b. The thermal emissions of the other three samples are presented in the supplementary material (Section 4). A detailed description of the measured emission normalization in the free-space system is given in the supplementary material (Section 5).

Figure 4 depicts the obtained emissivity and absorptivity, proving experimentally the generalization of Kirchhoff's law as presented by Eq.4 (ii). We note that the measured emissivity, after normalization is the hemispherical emissivity. This value is compared to the hemispherical absorptivity measured inside an integration sphere (see supplementary material, Section 6). The yellow line depicts the average hemispherical absorptivity at the spectral range (1450–1650 nm) of Erbium samples having different QEs. The QEs are measured at room temperature [25] and presented in the inset as a function of Erbium concentration. The blue line presents the hemispherical emissivity of each sample measured at the same spectral band as the absorptivity. We present the results obtained for 803K, while in the supplementary material (Section 7) we show the emissivity vs. temperature, which in our case is constant. Had the QE of the samples been zero, the blue and yellow lines would coincide, representing $\varepsilon = \alpha$. From figure 4 it is clear that the measured emissivity (blue line) coincides with the absorptivity multiplied by 1-QE of each sample (red line).

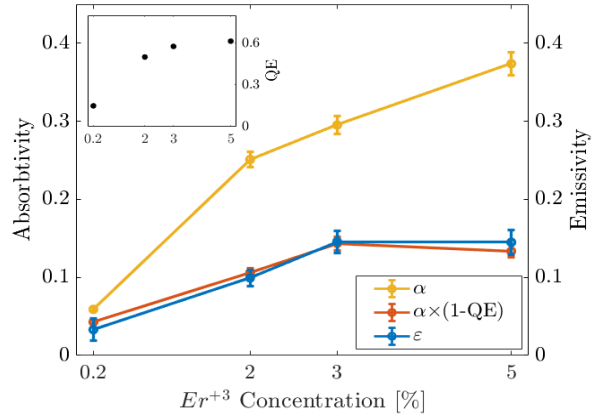

Fig. 4. Generalization of Kirchhoff's law for non-equilibrium radiation. Absorptivity of different QE samples, under optical excitation, measured at room temperature (yellow) and multiplied by 1-QE of each sample (red). The blue line is the measured emissivity of each sample. The inset presents the independently measured QE as a function of Er concentration.

Finally, we experimentally measure our hypothesis that the PL emission is the sum of a constant (QE) contribution at zero-kelvin and a temperature-dependent thermal emission. Figure 5 shows the measured difference between the PL and thermal emission of each sample vs. the temperature at ranges where the thermal emission can be obtained accurately (300K–750K). Clearly, all measurements show a constant QE contribution over thermal emission while both increase (500K–750K). The temperature gradient is noticeable only at high temperatures where it reaches the maximal gradient value of 50K, represented by the error bars. We note that in [4] the PL merges with thermal emission at temperatures above 1600K. This can be interpreted as quenching of the QE at high temperatures or as a result of the uncertainty in temperature that leads to high uncertainty in the flux.

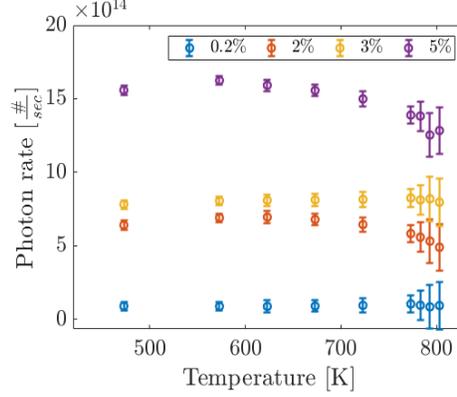

Fig. 5. Difference between PL and thermal emission rates in the measured temperature range.

## SUMMARY

To conclude, we theoretically, and experimentally demonstrate the prime equation for an emissivity as a material property in and out of equilibrium in the form of $\varepsilon = \alpha(1 - QE)$, relating emissivity, absorptivity, and QE. At equilibrium, this relation is reduced to Kirchhoff's law due to the balance of non-radiative (thermal) processes. Starting from the theoretical model describing a two-level system through the detailed balance approach, we derive this relation analytically and simulate it for a three-level system. Experimentally, we demonstrate the prime equation through direct measurement of its properties. We observe at high QE that emissivity is three-fold lower than absorptivity, with an excellent fit to the prime equation. Our work reveals the fundamental evolution of non-thermal emission with temperature. This new understanding is both important to fundamental science and for the engineering of lighting and energy devices.

**Funding.** European Union's Seventh Framework Program (H2020/2014-2020]) under grant agreement n° 638133-ERC-ThforPV.

**Acknowledgments.** The authors would like to thank Prof. Eli Yablonovich for fruitful discussions.

**Disclosures.** The authors declare no conflicts of interest.

**Data availability.** The data that support the findings of this study are available from the corresponding author upon reasonable request.

**Supplemental document.** See Supplement 1 for supporting content.